\documentclass[conference]{IEEEtran}
\usepackage{booktabs}
\usepackage{amsmath,amssymb,amsfonts}
\usepackage{algorithm}
\usepackage{algpseudocode}
\usepackage{graphicx}
\usepackage{xcolor}
\begin{document}
%\titleformat{\subsubsection}[hang]{\itshape}{\thesubsubsection)}{0.5em}{}
\title{Deep Reinforcement Learning for Scheduling in Cellular Networks}
\author{\IEEEauthorblockN{Jian ~Wang\IEEEauthorrefmark{1}, Chen ~Xu\IEEEauthorrefmark{1}, Yourui ~Huangfu\IEEEauthorrefmark{1}, Rong ~Li\IEEEauthorrefmark{1}, Yiqun ~Ge\IEEEauthorrefmark{2}, Jun ~Wang\IEEEauthorrefmark{1}}
\IEEEauthorblockA{\IEEEauthorrefmark{1}Hangzhou Research Center, Huawei Technologies, Hangzhou, China}
\IEEEauthorblockA{\IEEEauthorrefmark{2}Ottawa Research Center, Huawei Technologies, Ottawa, Canada}
Emails: \{wangjian23, xuchen14, huangfuyourui, lirongone.li, yiqun.ge, justin.wangjun\}@huawei.com
%\thanks{\vspace{-0.2cm}\hspace{-0.4cm}
%\hrule width 0.25\textwidth
%\vspace{0.15cm}
}

\maketitle
\begin{abstract}
Integrating artificial intelligence (AI) into wireless networks has drawn significant interest in both industry and academia. A common solution is to replace partial or even all modules in the conventional systems, which is often lack of efficiency and robustness due to their ignoring of expert knowledge. In this paper, we take deep reinforcement learning (DRL) based scheduling as an example to investigate how expert knowledge can help with AI module in cellular networks. A simulation platform, which has considered link adaption, feedback and other practical mechanisms, is developed to facilitate the investigation. Besides the traditional way, which is learning directly from the environment, for training DRL agent, we propose two novel methods, i.e., learning from a dual AI module and learning from the expert solution. The results show that, for the considering scheduling problem, DRL training procedure can be improved on both performance and convergence speed by involving the expert knowledge. Hence, instead of replacing conventional scheduling module in the system, adding a newly introduced AI module, which is capable to interact with the conventional module and provide more flexibility, is a more feasible solution.
\end{abstract}
\begin{IEEEkeywords}
artificial intelligence, cellular networks, deep reinforcement learning, scheduling, proportional fair
\end{IEEEkeywords}

\IEEEpeerreviewmaketitle

\section{Introduction}
\label{sec1}
Artificial intelligence (AI) technology has succeeded in injecting "intelligence" into a system to outperform humankind in many domains. Some have explored its applications into the wireless communication domain, as summarized in \cite{mao2018deep} and \cite{zhang2019deep}. Wireless network is a complex system consisting of several layers, each of which contains a functionality-specific module. Such a module is either represented or approximated by a mathematical expression. So far there are two major ways to adopt AI into the wireless network domain. The first is to use AI to replace end-to-end transceiver system completely, in which both transmitter and receiver are realized by neural networks (NNs). The second is to replace a part of the functionality-specific modules by NNs in the transceiver. Both suffer from a low efficiency and sub-optimal performance in missing of expert knowledge or comprehension about wireless environment. The presence of NN-based modules on a transceiver would not only undermine interoperability and stability, but also increases complexity. Therefore, we believe it more feasible for a AI module to work with traditional modules by mostly coordinating the working flows among these traditional modules, cooperating with them, or even replace them temporarily if needed. Instead of breaking into a well-defined system, an AI module could help this system to work better by its inherent learning and adapting capability.

To prove our belief, we choose scheduler as an example for a AI-enabled module in wireless network domain in this paper. A scheduler is the center of a cellular network system \cite{capozzi2013downlink}, decides how radio resources are allocated among users. A typical conventional scheduler considers channel conditions and QoS requirements to make allocation decisions based on some formulas. It is a deterministic way suffering from a low flexibility. To mitigate it, \cite{bertsekas1995dynamic} proposed a Markov Decision Process (MDP) solved by Dynamic Programming (DP). However, along with increasing size of the system, DP is too complex to be supported and the transition probabilities and states of the environment are too hard to known a priori. Naturally, deep reinforcement learning (DRL) offers an alternative to optimization over multiple dimensional data or signals. The DRL agent of the scheduler would observe the states of the environment, decides the actions, executes them into the environment, waits for the feedback reward, and transits to the next states. In this way, the agent learns from the environment and improves its decision-making strategy.

There have benn already several studies about scheduler based on DRL. The work in \cite{atallah2017deep} focusing on a Vehicle-to-Infrastructure (V2I) scenario uses DRL-based scheduler to extend the lifetime of the battery-powered Road-Side Units (RSUs) while promoting a safe environment that meets acceptable QoS levels. Chinchali et al. uses DRL-based scheduler to optimize the Internet of Things (IoT) traffic without impacting conventional real-time applications such as voice-calling and video \cite{chinchali2018cellular}. A joint user scheduling and content caching strategy based on DRL is studied in \cite{wei2018joint}, to decide whether to cache certain content and to select which small-cell base station (SBS) to serve certain UE simultaneously. In \cite{zhu2018new}, a DRL-based scheduler is designed to coordinate packet transmission from different buffers through multiple channels in a cognitive IoT network. An uplink scheduling problem is studied in \cite{chu2018reinforcement}, where DRL is used to choose $K$ from $N$ UEs to allow their uplink transmission to maximize throughput under power consumption consideration. These scenarios in these works are generally quite complex, because of multi-class services (e.g., IoT traffic and conventional traffic) or network devices (e.g., primary users and secondary users in cognitive networks, macro-cell BS and small-cell BS in heterogeneous networks). There is no ready-for-use scheduling algorithm in these scenarios, hence no one is able to provide a baseline for comparison. Although DRL agent was introduced therein, the capability, performance and efficiency thanks to DRL-based scheduler are difficult to study with no baseline. Moreover, except the one in \cite{chinchali2018cellular} that employed real field data for training and verification, others used loose assumptions to generate the training and test data samples. For instance, data rate is derived directly from Shannon's equation, without considering adaptive modulation and coding (AMC) and outer loop link adaption (OLLA). All undoubtedly eases the learning requirement but makes it even more dubious in practical systems.

In this paper, we go back to the case with the simplest settings in a cellular network and investigate scheduling based on DRL in the proposed AI-enabled network structure. We will firstly try a direct-learning method in which a DRL agent directly learns from the environment. Then, we propose two novel training methods: dual-learning and expert-learning. For the dual-learning method, two independent agents are trained alternatively and learns from each other. For expert-learning method, the PF algorithm is employed as expert knowledge to help with DRL agent training. By evaluating and comparing all the three methods, we could demonstrate the DRL's learning capability and how expert knowledge to help improve performance and accelerate training speed. To create an emulator environment as much close to true practical cellular network as possible, we developed a system simulator containing nearly all the link adaption, feedback and scheduling mechanisms used in real LTE networks.

The structure of this paper is as following. The preliminaries are first introduced in Section II. The network scenario, problem formulation and learning methodologies are described in Section III. Then, some trials are carried out in Section IV. Conclusions are made in Section V, where several future work directions are provided.

\section{Preliminaries}
\label{sec2}
\subsection{Proportional Fair (PF) Scheduling Algorithms}
\label{sec2.1}

%Typical scheduling algorithms can be classified into tree types. The simplest type does not consider channel conditions and QoS requirements, hence, it cannot perform adaptive adjustment according to channel changing and provide differential QoS provisioning. Round-Robin (RR) is the most famous algorithm in this category. Radio resources are allocated among UEs with equal probability, no matter how good or bad channels they experience or how high or low their priorities are. By using the CQI feedbacks, the second type of scheduling algorithms can be designed based on not only the historic information but also the prediction of future channel quality. Maximum Carrier-to-Interference (Max C/I) algorithm aims at maximize the overall throughput by allocating the resources to UEs with the best channel conditions. This strategy sacrifices the fairness among UEs, since those with poor channel conditions may always be kept out of the scheduling decision. The third type of scheduling algorithms takes both channel conditions and QoS requirements into account. There always exists UEs with different QoS requirements in the practical systems. The QoS requirements can be represented by a set of QoS parameters, which are then input into the scheduling algorithms together with CQI feedback. QoS, in form of such as data rate and delay, can then be guaranteed by taking them into consideration when calculating scheduling metrics for each UEs.

Proportional Fair (PF) is among the most widely used scheduling algorithms. It provides a tradeoff between system overall throughput and fairness among UEs. Therefore, we set PF as the baseline in this paper. As a preliminary, we brief PF algorithm in this section as follows. Consider a scheduling scenario in which a total amount of resource (system capacity) $C$ is allocated among $N$ UEs in the set $\cal{N}$. Each UE has a data rate $x_n, n \in \cal{N}$. A set of rates $\left\{ {x_n ,n \in \cal{N}} \right\}$ is proportionally fair if feasible (i.e., $x_n \geq 0$ and sum of $x_n$ is no larger than $C$) and if for any other feasible set $\left\{ {x_n^* ,n \in \cal{N}} \right\}$, the aggregate of proportional changes is zero or negative \cite{kelly1997charging}:
\begin{equation}\label{eq:pfstate}
\sum\limits_{n \in \cal{N}} {{{x_n^*  - x_n } \over {x_n }}}  \le 0
\end{equation}

F. Kelly formulates it into an utility maximization problem:
\begin{equation}\label{eq:ump}
\begin{split}
  \max \quad &\sum\limits_{n \in \cal{N}} {U_n \left( {x_n } \right)} \\
  {\rm{s}}{\rm{.t}}{\rm{.}}\quad & \;\;x_n \;{\rm{is}}\;{\rm{feasible}}
\end{split}
\end{equation}
where it is ``feasible'' that all the data rates should be no smaller than zero and the total summation of data rates should be no larger than the system capacity. When the utility function $U_n$ is logarithm function, the solution of problem (\ref{eq:ump}) has the unique vector of rates, which is named as proportionally fair \cite{kelly1997charging}. Therefore, the PF scheduling algorithm is optimal in term of maximizing the sum of logarithmic rate.

Kelly provides a general principle about proportional fairness. Later, \cite{tse2001multiuser} adopts it into the scenario of wireless networks with a single carrier. The only carrier (or say channel) should be allocated to the UE with the largest metric of $I_n /T_n $, where $I_n$ is the instantaneous throughput estimated from the updated channel condition and $T_n$ is the average throughput within a past window for the $n$-th UE, repectively. This metric indicates that those UEs with higher instantaneous throughput (i.e., better channel condition) should have higher priority to access the channel resource, to improve the overall throughput of the system, while at the same time, whose with smaller historic throughput should be given more chance to access the channel, to guarantee some degree of fairness. To summarize, for a single-carrier system, to achieve proportional fair, UE should be chosen according to
\begin{equation}\label{eq:pf}
i = \mathop {\arg \max }\limits_{n \in \cal{N}} {{I_n } \over {T_n }}
\end{equation}
where, the average throughput is updated according to
\begin{equation}\label{eq:ave_rate}
T_{n}\left ( t \right ) = \frac{W-1}{W}T_{n}\left ( t-1 \right ) + \frac{1}{W}I_{n}\left ( t \right )
\end{equation}
with $W$ as the window size for averaging.

%For multi-carrier systems, \cite{kim2005proportional} provides the method to achieve proportional fairness, and a greedy way is elaborate in \cite{sun2006reduced}.

%While, it has been proven that the aforementioned simple method can achieve proportional fairness in a single-carrier system, it is not that simple to extend this result to a multi-carrier system. \cite{kim2005proportional} shows that an algorithm indeed exists for multi-carrier systems that can achieve proportional fairness. Consider a total set of carriers $K$ is allocated among UEs in the set $S$, a subset of UEs $E$ are chosen from $S$ by scheduler to use the carriers, then this subset should be chosen according to
%\begin{equation}\label{eq:mcpf}
%P = \mathop {\arg \max }\limits_{E \subset S} \prod\limits_{s \in E} {\left( {1 + {{\sum\limits_{k \in K_s } {I_{s,k} } } \over {\left( {W - 1} \right)T_s }}} \right)}
%\end{equation}
%where, the set $K_s$ contains all the carriers allocated to UE $s$, and $I_{s,k}$ is the instantaneous throughput of UE $s$ on carrier $k$. The calculation of Equation \ref{eq:mcpf} is not that efficient, hence a greedy algorithm is usually used in practice, where UEs are chosen carrier by carrier\cite{sun2006reduced}. For an unallocated carrier $k^*$, one UE is chosen according to
%\begin{equation}\label{eq:mcpf_greedy}
%i = \mathop {\arg \max }\limits_{s \in S} {{I_{s,k^* } } \over {\left( {W - 1} \right)T_s  + \sum\limits_{k \in K_s } {I_{s,k} } }}
%\end{equation}
%Repeating this UE selection step for each unallocated carrier one by one.

\subsection{Deep Reinforcement Learning}
\label{sec2.2}
An MDP is typically defined by state space $\cal S$, action set $\cal A$, reward function $\cal R$ and transition probability $\cal P$. The target is to find a policy $\pi$ that maximizes the expected (discounted) reward. Reinforcement learning (RL) can be utilized to solve MDP if the full knowledge is unavailable, e.g., $\cal R$ and/or $\cal P$ are unknown. The RL agent learns from the interactions with the environment gradually. There are such classical RL algorithms as Q-learning, Policy Gradient (PG), Actor Critic, etc. They are useful to solve MDP with small scales of states. However, out practical MDP has usually large-scale states and complicated state transitions, so that the computational complexity of classical RL algorithms quickly becomes unaffordable. As a result, DNNs are introduced to improve both the learning performance and training speed.

Deep Q-Network (DQN) is a typical RL algorithm that uses a DNN to approximate Q value. Quadruples of $(S^t, A^t, R^t, S^{t+1})$ that represents current state, current action, current reward, and next state are stored as experience, from which a mini-batch would randomly samples a number of the quadruples (experience replay) to train the DNNs. In this way, both new and old quadruples have chance to be included in the mini-batch. A typical DQN contains two architecture-identical DNNs: target network and evaluation network. The target network is fixed for several training steps; whereas the evaluation network is updated every episode. After a given period, the target network would be updated by the evaluation one. $\epsilon$-greedy algorithm is used for some probability of exploration.

Although DQN (value-based method) is good at making discrete decisions, systems in many applications have continuous state/action spaces. Policy-based methods are more efficient in these problems because they directly optimize on the policy $\pi$. In addition, the actor-critic algorithm learns both a value function and a policy to reduce the gradient variance of PG. Deep deterministic policy gradient (DDPG) \cite{lillicrap2015continuous} is introduced as the extension of DQN and deterministic policy gradient, which is an off-policy actor-critic algorithm suitable for high-dimension continuous problems. DDPG employs experience replay and soft target network similar to DQN in order to improve the training stability.

\section{Network Scenario and Simulation Platform}
\label{sec3}
\subsection{Network Scenario and Simulation Platform}
\label{sec3.1}
In this paper, we consider a single cell cellular network following LTE standard. The BS is equipped with an AI scheduler. This AI-enabled scheduler doesn't break into the established conventional modules but schedules them.

It has been mentioned in Section \ref{sec2.1} that, in a single-carrier system, PF algorithm based on Equation (\ref{eq:pf}) can achieve proportional fairness. Hence, we will investigate this single-carrier system by assuming there is only one resource block group (RBG) containing all the resource blocks (RBs) in the system. Then, DRL algorithms (DDPG) are installed in the AI module to see if and how they can cooperate with the scheduling module to achieve the optimum.

\begin{figure}[t]
	\centering
	\includegraphics[width=.9\columnwidth]{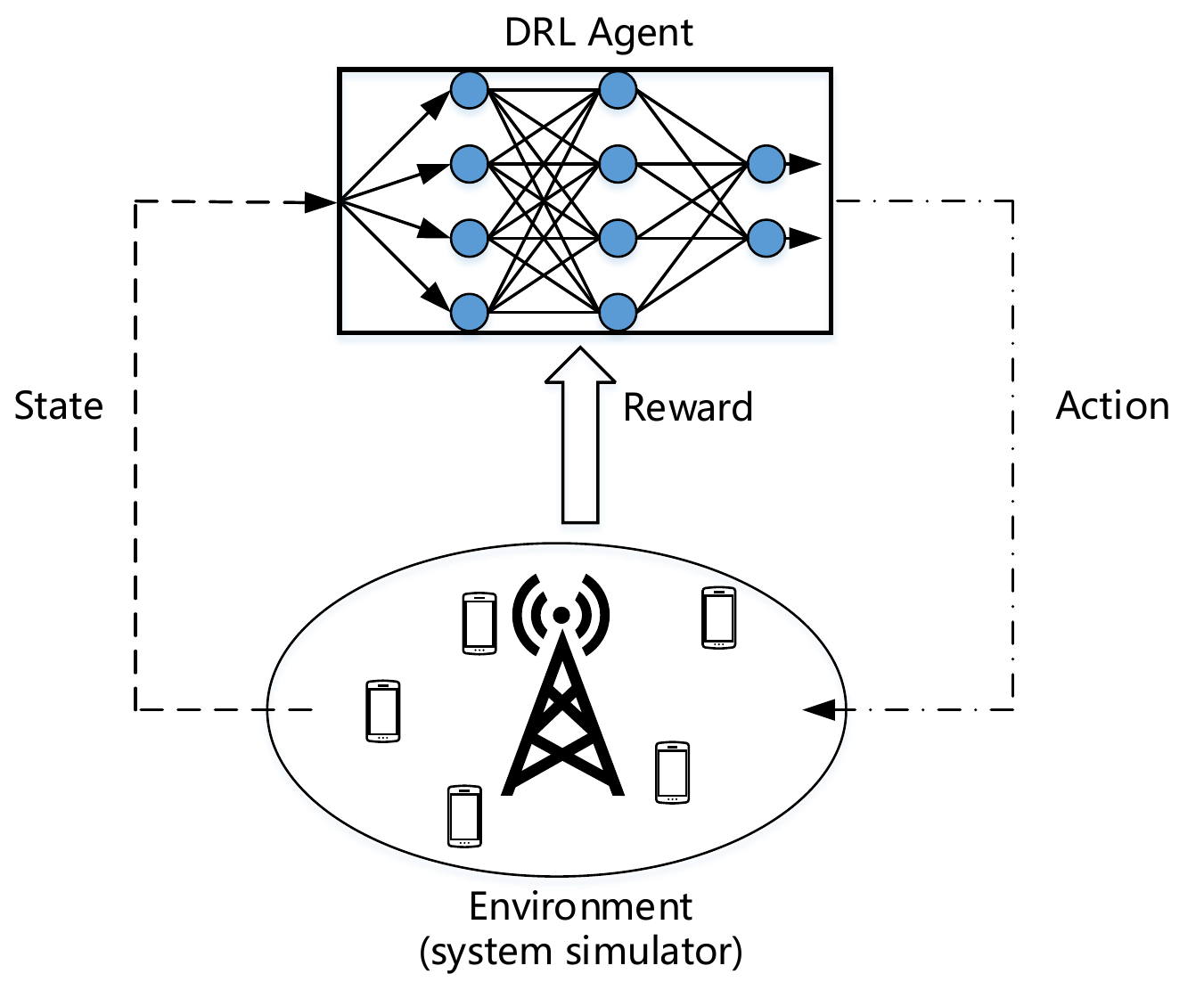}
	\caption{Simulation platform.}
	\label{fig:interaction}
\end{figure}

As shown in Fig.~\ref{fig:interaction}, the simulation platform contains a DRL agent realized in Tensorflow \cite{tensorflow2015-whitepaper} and a system simulator working as the environment. The system simulator provides the interfaces for the agent to acquire states and rewards, and execute actions.

%DDPG is employed as the DRL algorithm. Both the actor and the critic in DDPG framework are realized through networks with two hidden layers. The number of neurons for each hidden layer is set to 16 times of the number of UEs in the network. The dimensions of input and output layers are set according to those of system states and actions, respectively.

\begin{figure}[t]
	\centering
	\includegraphics[width=0.8\columnwidth]{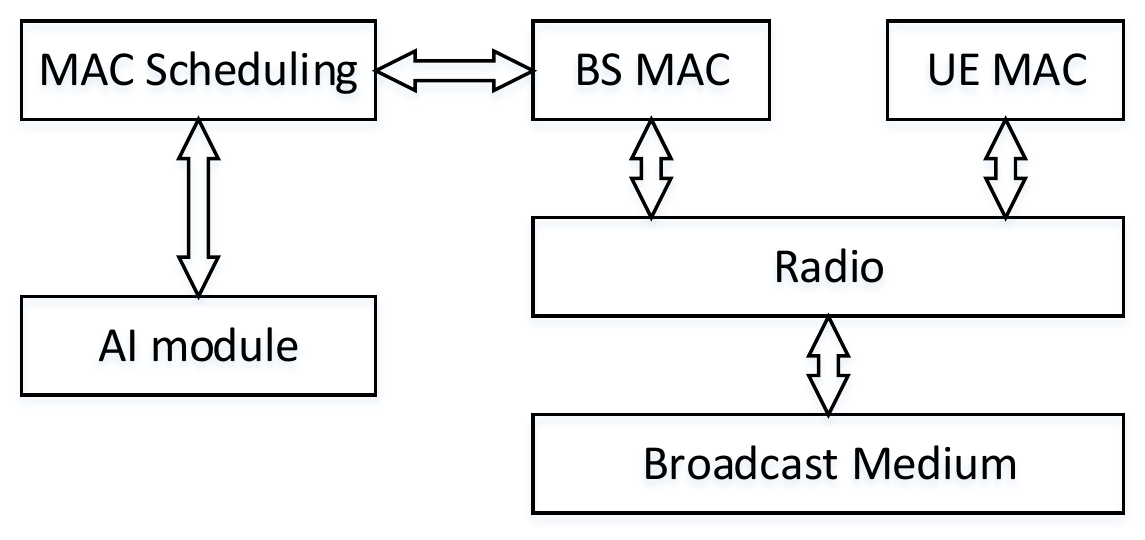}
	\caption{Functional structure of the system simulator.}
	\label{fig:syssimulator}
\end{figure}

The structure of the system simulator is shown in Fig.~\ref{fig:syssimulator}. The BS MAC and UE MAC module are responsible for transferring data packets as well as generating and handling control channel signals such as HARQ and CSI reporting. Meanwhile, the BS MAC relies on the scheduling module to allocate resources to UEs. The Radio module has two main functions: to check if the data packets can be decoded successfully or not, and to calculate the feedback quantities such as SINR and rank. The Broadcast Medium module keeps tracking on antenna sets and their channels. The module mainly concerned by this paper is the MAC scheduler that focuses on making resource allocation decision based on the channel condition and historic throughput information. An AI-enabled module is embedded in the system to help the scheduling procedure. This system simulator realizes nearly all the link adaption and feedback function in LTE. The AMC function follows exactly the LTE standard, that modulation and coding schemes (MCSs) are chosen from LTE MCS table with a target block error ratio (BLER). A fixed step outer loop link adaption is also employed to compensate the feedback error due to the imperfection of feedback channel.

\begin{figure}[t]
	\centering
	\includegraphics[width=\columnwidth]{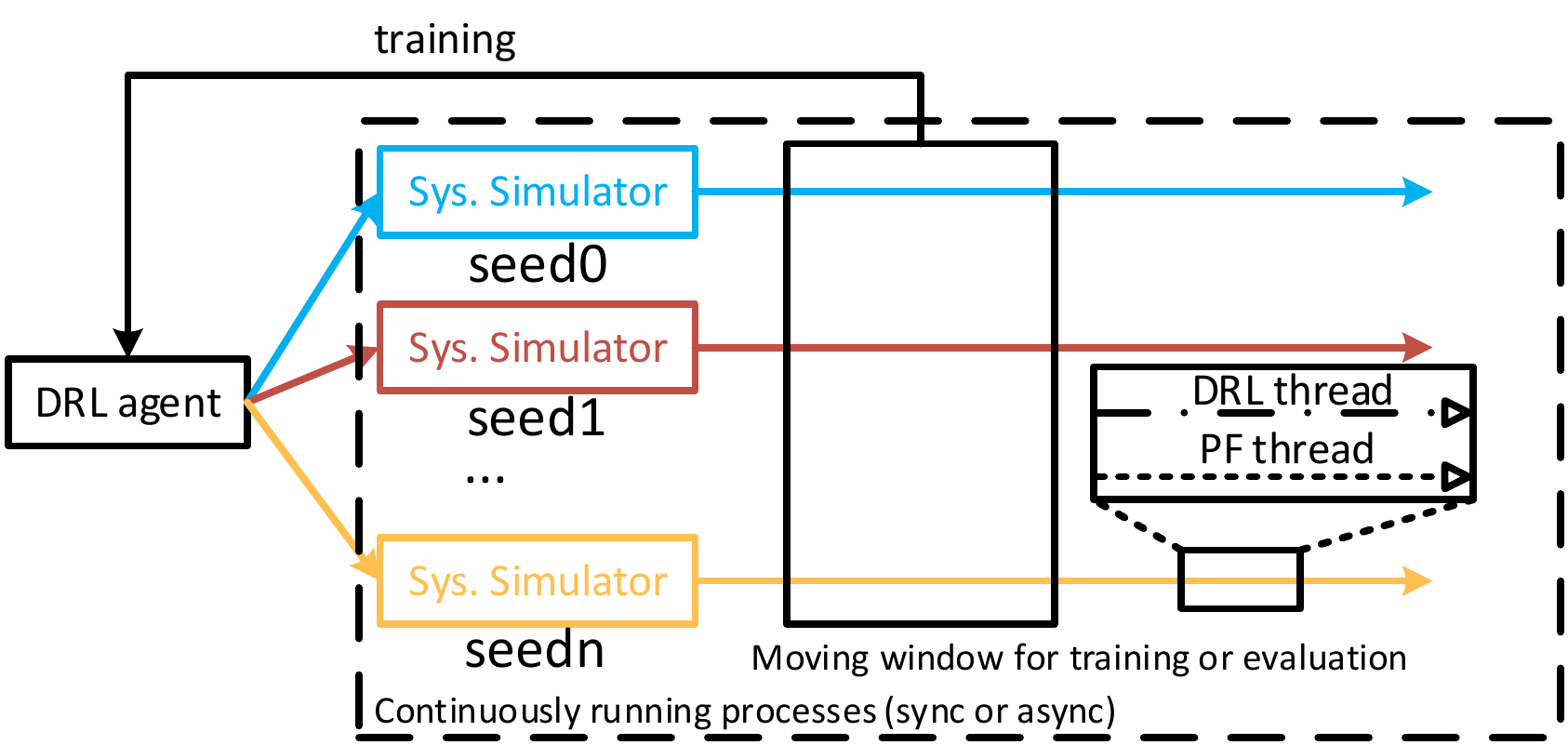}
	\caption{Simulation framework.}
	\label{fig:framework}
\end{figure}

As shown in Fig.~\ref{fig:framework}, multiple simulations with different UE positions and data generation seeds are started simultaneously, so that the experience buffer contains the samples with different UE deployments. These various samples help DRL agent to generalize, i.e., make good decisions in different situations. For each system simulator, two threads are initialized, one interacts with the DRL agent, and the other executes PF scheduling algorithm for performance comparison. Moving windows is used to restrict the time scope of training and evaluation.

\subsection{Problem Formulation}
\label{sec3.2}
The MDP of this scenario is elaborated as follows.
\begin{itemize}
  \item State: the instantaneous rate $I_n(t)$ and average rate $T_n(t)$ for each UE $n \in \cal{N}$ are contained in the state $S^{t}$. Reminding that the update of average rate as shown in Equation (\ref{eq:ave_rate}) involves the influence of the previous action through the term $I_n(t)$, thus the transition from state $S^{t}\doteq \left ( I_n(t), T_n(t)\right )$ to state $S^{t+1}$ is Markovian.
%      Later on, we also study the case where the buffer condition and delay tolerance are taken into account. Finite-length buffer will lead to packet loss due to overflow, while packets which have expired will also be dropped before transmitted. These two kinds of packet loss influence with each other, because smaller buffer which induces higher overflow will keep the delay of packets in the buffer at a low level, while lower delay tolerance, through results in more expired packets dropping, helps to prevent the buffer from flooding. When considering these two factors, the state of the MDP becomes $S^{t}\doteq \left ( I^{t}, T^t, B^t, D^t\right )$, where $B^t$ represents the buffer condition (number of packets in the buffer with equal size for simplicity), and $D^t$ represents the already waiting time of the head of line packet (which the buffer realized by a drop-tail queue).
  \item Action: Since a single-carrier scheduling scenario is considered, the action is to decide which UE is scheduled to occupy the only RBG in each scheduling period (i.e., Transmission Time Interval, TTI in LTE). Similar to PF algorithm, for DRL, the action output $A^t$ is designed to be metrics for each UE. The BS can choose the UE with the largest metric as the scheduled UE in each TTI.
  \item Reward: The reward design is fundamental in DRL, because it has impacts on both system performance and training speed. In this paper, both the total throughput and UE fairness are considered into the reward. To quantify the UE fairness, Jain's fairness index (JFI) \cite{jain1984quantitative} is used. The calculation of JFI is shown in Equation (\ref{eq:fairness}), where $V_n, n\in \cal{N}$ represents the received average throughput of the $n$-th UE. The details of reward calculation can be explained from learning methodology perspective elaborated in Section \ref{sec3.3}.
\begin{equation}\label{eq:fairness}
fairness = {{\left[ {\sum\limits_{n = 1}^N {V_n } } \right]}^2 \mathord{\left/
 {\vphantom {{\left[ {\sum\limits_{n = 1}^N {V_n } } \right]} {\left[ {N\sum\limits_{n = 1}^N {V_n^2 } } \right]}}} \right.
 \kern-\nulldelimiterspace} {\left[ {N\sum\limits_{n = 1}^N {V_n^2 } } \right]}}
\end{equation}
%      With taking buffer and delay into account, packet loss should also be reflected in the reward calculation.

\end{itemize}

\subsection{Learning Methodology}
\label{sec3.3}
\begin{figure*}[t]
	\centering
	\includegraphics[width=2\columnwidth]{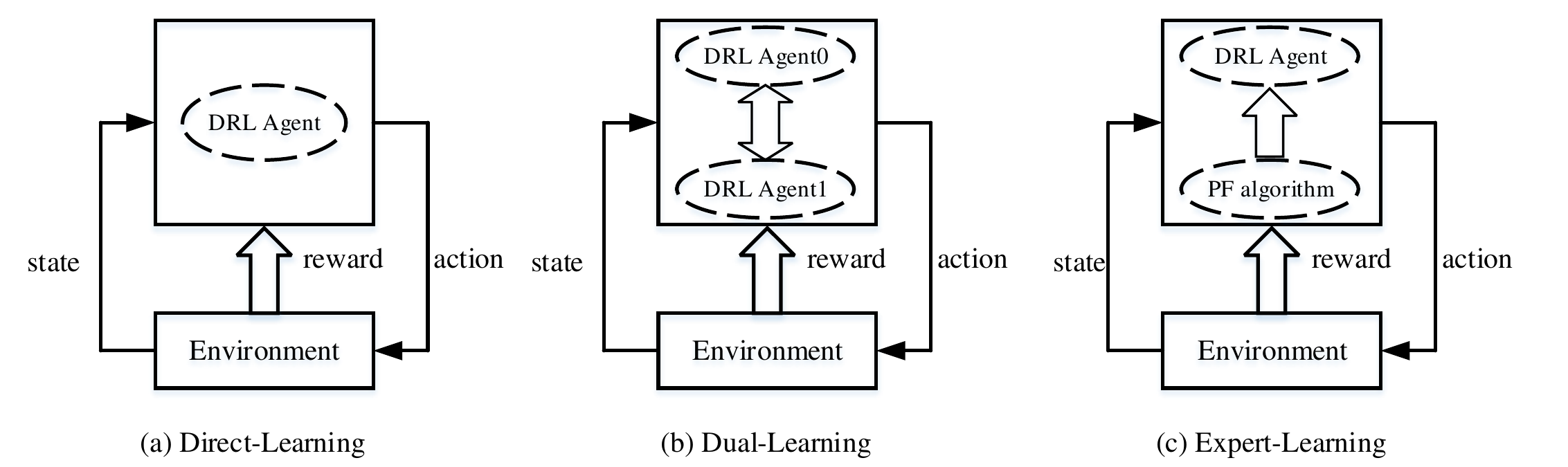}
	\caption{Learning methodologies.}
	\label{fig:learning}
\end{figure*}
In this section, three methods to calculate reward are elaborated. In terms of their learning procedures, we classify them into three different learning methodologies for DRL as shown in Fig. \ref{fig:learning}.

\subsubsection{Direct-Learning}
\label{sec3.3.1}
\begin{equation}\label{eq:linear_reward}
reward = \alpha \times throughput + \beta \times fairness
\end{equation}

The first learning methodology is direct-learning, where the DRL agent learn directly from the environment without considering the expert knowledge. Taking throughput and fairness into consideration, Equation (\ref{eq:linear_reward}) is the most straightforward reward function. Here, the $throughput$ in the reward function is the received instantaneous throughput, hence the UE with the best channel condition tends to be chosen. The $fairness$ is calculated through Equation (\ref{eq:fairness}). $\alpha$ and $\beta$ are weighting factors for throughput and fairness, respectively. Apparently, set $\alpha = 1$ and $\beta = 0$ will lead to a performance similar to Max C/I. Without expert knowledge in this reward function, the DRL agent is trained to learning directly from the environment and improve both throughput and fairness simultaneously.

\subsubsection{Dual-Learning}
\label{sec3.3.2}
%Generally, a non-stationary environment should be modeled as hidden Markov model (HMM) and use partially observable Markov decision process (POMDP) algorithms.
\begin{algorithm}[h]
	\caption{Dual-learning framework}
	\label{alg:self_learning}
	\begin{algorithmic}
		\State Initialize two independent agents
		\State Initialize two environments with the same seed
		\While {not convergent}
		\State \Call{Train}{$agent_0, agent_1$}
		\State \Call{Train}{$agent_1, agent_0$}
		\EndWhile
		
		\Function{Train}{$agent_i, agent_j$}
		\State Freeze parameters of $agent_j$
		\State $agent_i$ and $agent_j$ interacts with environments
		\State Calculate reward for $agent_i$ using the lookup table
		\State Update parameters of $agent_i$
		\EndFunction
	\end{algorithmic}
\end{algorithm}

\begin{table}[h]
	\centering
	\caption{Dual-Learning Reward Look-up Table}
	\begin{tabular}{lcccccc}
		\toprule
							&				&				 &	Agent0	&		& \\
		\midrule
		Throughput	 		&				&$>$			 &$>$		&$\le$	&$\le$\\
		JFI 				&Agent1			&$>$			 &$\le$		&$>$	&$\le$\\
		Reward				&	 			&$\alpha + \beta$&$\alpha$	&$\beta$&0\\
		\bottomrule
	\end{tabular}
	\label{tab:self_reward_table}
\end{table}

In this subsection, we propose an alternative framework of training DRL agents, i.e., dual-learning, in the specific scheduling problem. As described in Algorithm~\ref{alg:self_learning}, two independent agents are randomly initialized to interact with the same environment. By training them alternatively, two agents can help each other set up a baseline and then learn from it, eventually achieving global optimum.

Reward calculation for this framework is straightforward: reward increases by 1 if certain performance metric is better than that of the other agent, or the increment is 0. The purpose of the reward is that the DRL agent is encouraged to make better decisions than the opponent agent. To enable flexibility of adjusting contribution weight of throughput and fairness, $\alpha$ and $\beta$ are also introduced. Thus the reward can be designed into a look-up table, as seen in Table~\ref{tab:self_reward_table}. This dual-learning procedure involves no expert knowledge from conventional scheduling algorithms.

\subsubsection{Expert-Learning}
\label{sec3.3.3}
\begin{table*}[t]
	\centering
	\caption{Expert-Learning Reward Look-up Table}
	\begin{tabular}{lccccccccccc}
		\toprule
							&			&				&			&		&   &&PF&&&            \\
		\midrule
		Throughput	 		&			&$>$			&$>$		&$<$	&$<$&$>$&$=$&$=$&$=$&$<$            \\
		JFI 				&RL			&$>$			&$<$		&$>$	&$<$&$=$&$>$&$=$&$<$&$=$            \\
		Reward				&	 		&$\alpha + \beta$&$\alpha$	&$\beta$&0  &$\alpha + 0.5\beta$&$0.5\alpha + \beta$&$0.5(\alpha + \beta)$&$0.5\alpha$&$0.5\beta$            \\
		\bottomrule
	\end{tabular}
	\label{tab:pf_reward_table}
\end{table*}

As the PF algorithm given by Equation (\ref{eq:pf}) is proven to be a good tradeoff between throughput and fairness, we can take advantage of it and speed up the training of DRL agent. For this expert-learning method, the reward rule of dual-learning method is reused, but one opponent agent is replaced with the PF algorithm. Hence we have a similar reward table Table~\ref{tab:pf_reward_table}. Note that the reward will increase by 0.5 if certain metric is equal to PF algorithm because of PF's optimality.

\section{Trials and Discussions}
\label{sec4}
In this section, we present the simulation results. A 10MHz bandwidth single cell LTE network is considered with 5 UEs receiving full-buffer traffic packets from the BS. The NNs used in the DRL agent are fully connected ones with 2 hidden layers, each of which contains 320 neurons. ReLU function is used as the activation function for all the hidden layers. Softmax is used at the output layer of the actor network and no activation function is used for the output layer of the critic network. The DRL agents are trained with 28 independently but simultaneously running system simulators for better convergence and generalization ability. We take PF as baseline and compare DRL performance with it every 50 training updates during the training process. The normalized performance difference between DRL and PF algorithm is elaborated in the y-axes in the figures below, where a positive value means DRL algorithm performs better than the baseline. The x-axes present the counts of training updates.

\subsection{Direct-Learning}
\label{sec4.1}
\begin{figure}[t]
	\centering
	\includegraphics[width=0.9\columnwidth]{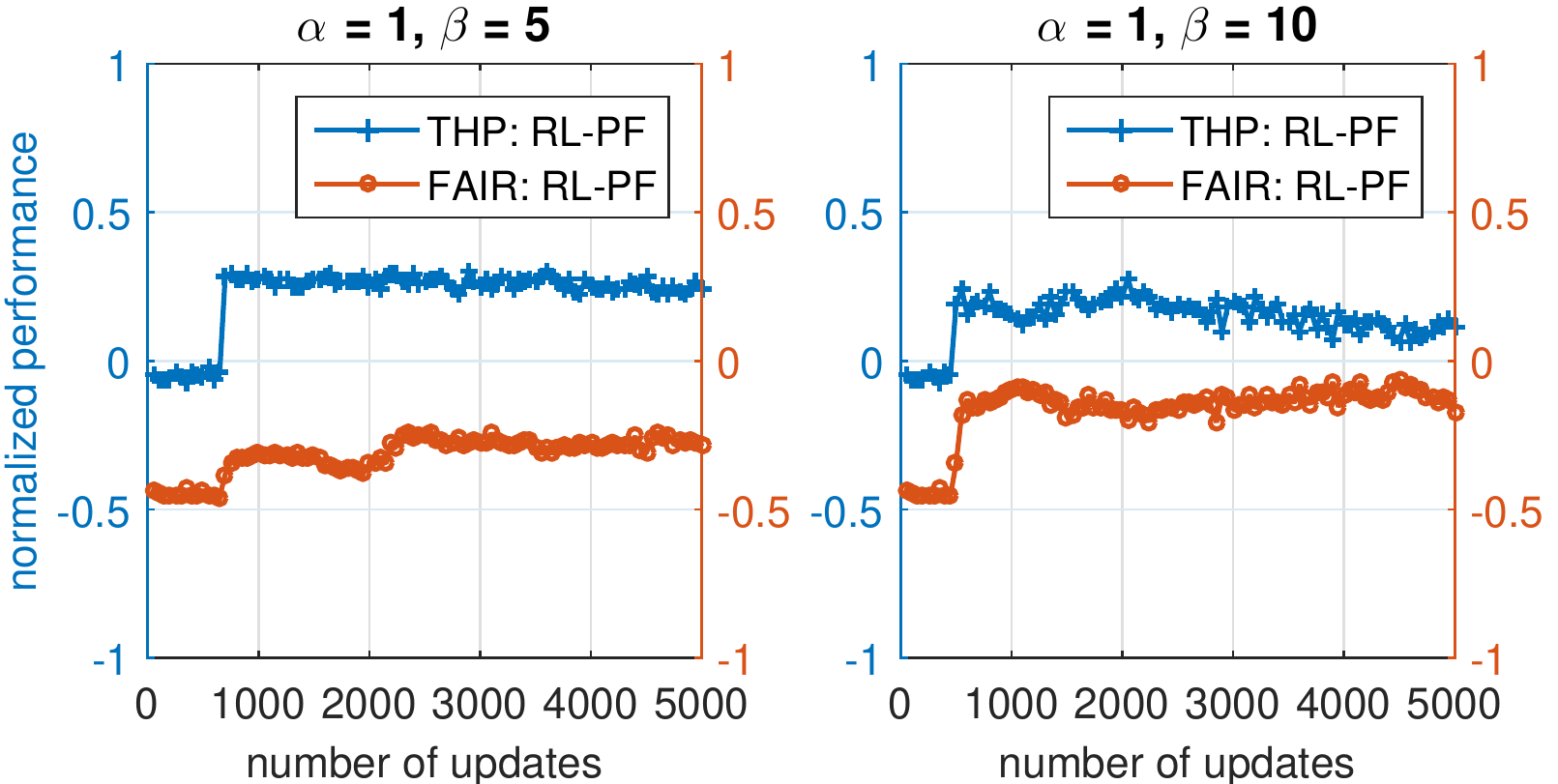}
	\caption{Comparison between direct-learning RL agent and PF algorithm.}
	\label{fig:direct_learn}
\end{figure}

As shown in Fig.~\ref{fig:direct_learn}, the DRL algorithm converges quickly in $\sim$1000 updates, and then, nearly no obvious improvement can be obtained. The results show that DRL algorithm can achieve higher throughput with the sacrifice of fairness. To make fairness take higher weight in the reward function, we tried to adjust the weighting factors, e.g., from $\beta / \alpha = 5$ to $\beta / \alpha = 10$. The adjustment on $\alpha$ and $\beta$ does help with the tradeoff between throughput and fairness, but PF state is still difficult to approach according to our simulations. The reason is that the wireless environment is non-stationary, that is, the reward may be different in training even if with the same state and action, thus the agent is hard to converge or quickly falls into local optimum.

\subsection{Dual-Learning}
\label{sec4.2}
For the dual-learning procedure, we choose $\alpha = 0.85$ and $\beta = 1.05$ in this experiment.

\begin{figure}[h]
	\centering
	\includegraphics[width=0.9\columnwidth]{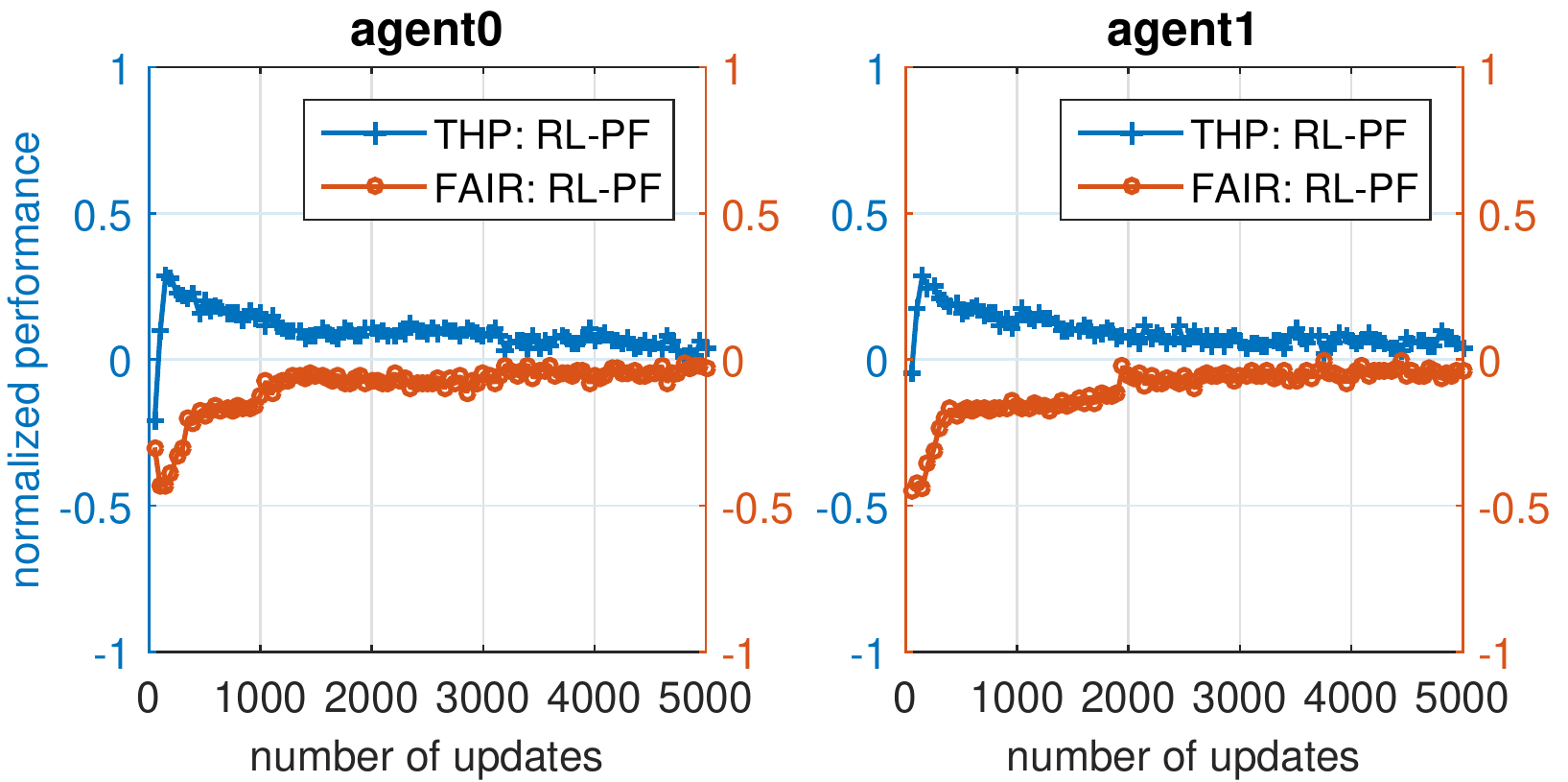}
	\caption{Comparison between dual-learning RL agent and PF algorithm.}
	\label{fig:dual_learn}
\end{figure}

As shown in Fig.~\ref{fig:dual_learn}, after $\sim$2000 updates for each agent, two agents both converge to near-optimal. This method helps the DRL agent continuously make progress towards the global optimal solution. The drawback is that still no expert knowledge is used, hence the convergence speed is slow.

\subsection{Expert-Learning}
\label{sec4.3}

\begin{figure}[t]
	\centering
	\includegraphics[width=.9\columnwidth]{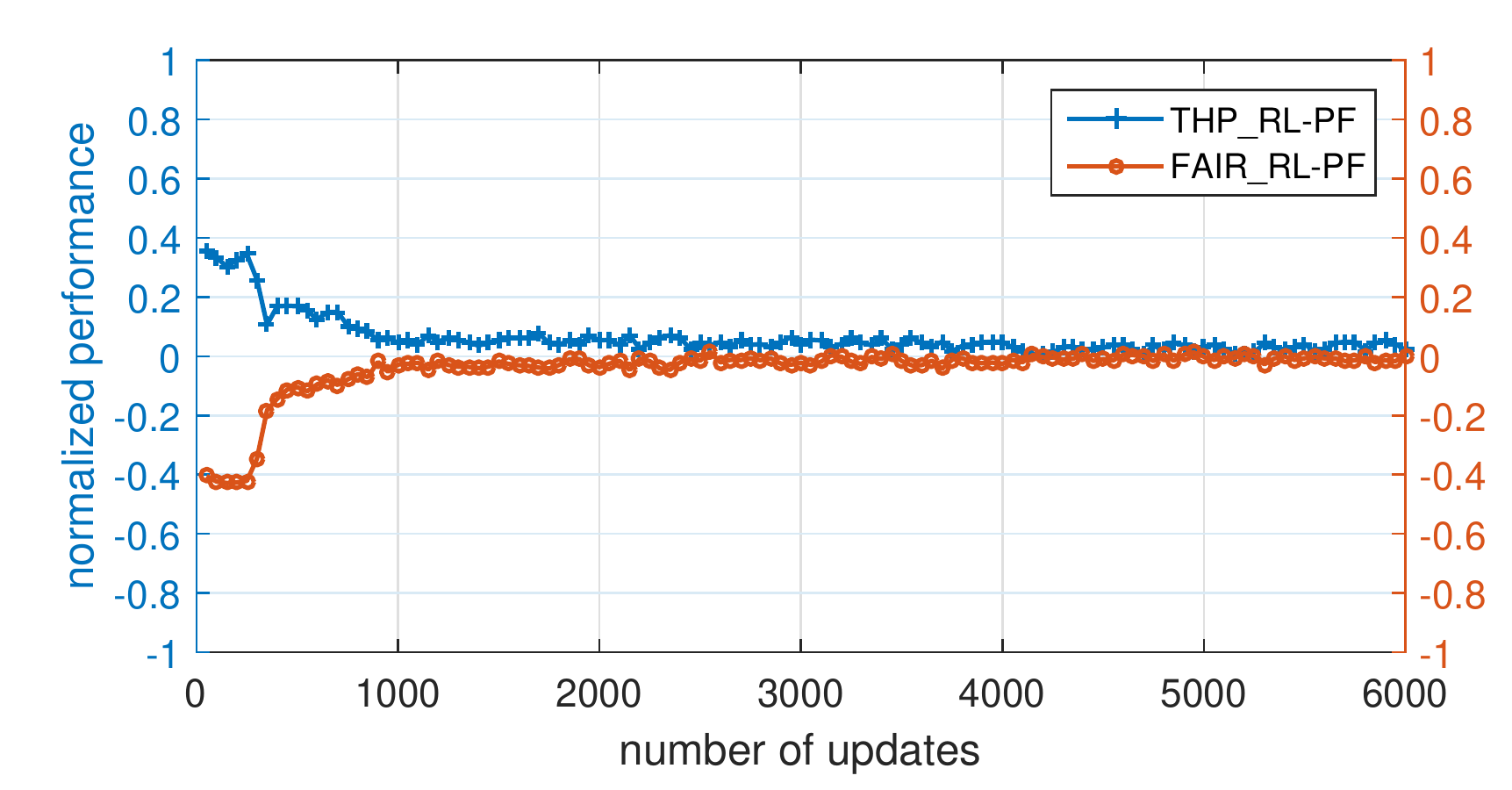}
	\caption{Comparison between expert-learning RL agent and PF algorithm.}
	\label{fig:pf_learning}
\end{figure}

We choose $\alpha = 0.9$ and $\beta = 1.15$ in this experiment. Fig.~\ref{fig:pf_learning} records the throughput and JFI differences between DRL agent and PF algorithm. It can be seen that after $\sim$1000 updates, DRL agent converges to nearly optimal, and has successfully learnt from PF after $\sim$4000 updates, achieving optimum.

\subsection{Discussions}
\label{sec4.4}
To conclude the above subsections, it is found for the investigated scheduling problem that:
\begin{enumerate}
  \item DRL agent can learn directly from the environment. By setting different weights for throughput and fairness in the reward function, the algorithm can converge to different points. However, it still has a risk to fall into local optimal points.
  \item Dual-learning can help with getting rid of falling into local optimum, but the training speed is relatively slow due to ignoring expert knowledge.
  \item If good expert knowledge exists, learning from it can speed up the training procedure.
\end{enumerate}

The results in this section suggest that instead of replacing the conventional PF scheduler by an AI scheduler, add one is more feasible. This AI scheduler can learn online from the PF scheduler. When number of UEs in the system becomes large and the requirements of them becomes different, PF scheduling is slow and inefficient. The AI scheduler can help to make flexible decisions at one shot. Meanwhile, although the generalization capability of NNs is proven to be strong in some cases, it is more robust to keep the adjustment of NNs according to the changing environment. The AI scheduler can be trained online when the PF scheduler is working in the system. The switch of PF and AI scheduler can be designed so that system performance is optimized and not influenced while the AI scheduler is training.

\section{Conclusions and Future Work}
\label{sec5}
In this paper, we investigate how DRL can help with solving the scheduling problem in cellular networks. Three learning methodologies are introduced while training the DRL agent. Results show that, in the scenario of scheduling, learning from the expert knowledge can provide the best performance and the most efficient training. So, instead of replacing the conventional scheduling module, add an AI scheduler is more feasible.

For the future, more state and performance parameters can be considered. For instance, considering a finite-length buffer with non-full buffer traffic and a delay tolerance for each packet, the packet loss due to buffer overflow and time expiration will surely influence the system performance. The DRL agent should take buffer and delay condition as parts of the state, and packet loss ratio as a component of reward calculation. 

%\section*{Acknowledgment}
%This work was supported in part by the National Science Foundation of China under Grants 60972058, 60802012 and 61001098, and National Key Basic Research Program (973 Program) under Grant 2009CB320405.

{\footnotesize
\bibliographystyle{IEEEtran}
\bibliography{rlscheduling}

\begin{thebibliography}{10}
\providecommand{\url}[1]{#1}
\csname url@samestyle\endcsname
\providecommand{\newblock}{\relax}
\providecommand{\bibinfo}[2]{#2}
\providecommand{\BIBentrySTDinterwordspacing}{\spaceskip=0pt\relax}
\providecommand{\BIBentryALTinterwordstretchfactor}{4}
\providecommand{\BIBentryALTinterwordspacing}{\spaceskip=\fontdimen2\font plus
\BIBentryALTinterwordstretchfactor\fontdimen3\font minus
  \fontdimen4\font\relax}
\providecommand{\BIBforeignlanguage}[2]{{%
\expandafter\ifx\csname l@#1\endcsname\relax
\typeout{** WARNING: IEEEtran.bst: No hyphenation pattern has been}%
\typeout{** loaded for the language `#1'. Using the pattern for}%
\typeout{** the default language instead.}%
\else
\language=\csname l@#1\endcsname
\fi
#2}}
\providecommand{\BIBdecl}{\relax}
\BIBdecl

\bibitem{mao2018deep}
Q.~Mao, F.~Hu, and Q.~Hao, ``Deep learning for intelligent wireless networks: A
  comprehensive survey,'' \emph{IEEE Communications Surveys \& Tutorials},
  vol.~20, no.~4, pp. 2595--2621, 2018.

\bibitem{zhang2019deep}
C.~Zhang, P.~Patras, and H.~Haddadi, ``Deep learning in mobile and wireless
  networking: A survey,'' \emph{IEEE Communications Surveys \& Tutorials},
  2019.

\bibitem{capozzi2013downlink}
F.~Capozzi, G.~Piro, L.~A. Grieco, G.~Boggia, and P.~Camarda, ``Downlink packet
  scheduling in {LTE} cellular networks: Key design issues and a survey,''
  \emph{IEEE Communications Surveys \& Tutorials}, vol.~15, no.~2, pp.
  678--700, 2013.

\bibitem{bertsekas1995dynamic}
D.~P. Bertsekas, D.~P. Bertsekas, D.~P. Bertsekas, and D.~P. Bertsekas,
  \emph{Dynamic programming and optimal control}.\hskip 1em plus 0.5em minus
  0.4em\relax Athena scientific Belmont, MA, 1995, vol.~1, no.~2.

\bibitem{atallah2017deep}
R.~Atallah, C.~Assi, and M.~Khabbaz, ``Deep reinforcement learning-based
  scheduling for roadside communication networks,'' in \emph{2017 15th
  International Symposium on Modeling and Optimization in Mobile, Ad Hoc, and
  Wireless Networks (WiOpt)}.\hskip 1em plus 0.5em minus 0.4em\relax IEEE,
  2017, pp. 1--8.

\bibitem{chinchali2018cellular}
S.~Chinchali, P.~Hu, T.~Chu, M.~Sharma, M.~Bansal, R.~Misra, M.~Pavone, and
  S.~Katti, ``Cellular network traffic scheduling with deep reinforcement
  learning,'' in \emph{Thirty-Second AAAI Conference on Artificial
  Intelligence}, 2018.

\bibitem{wei2018joint}
Y.~Wei, Z.~Zhang, F.~R. Yu, and Z.~Han, ``Joint user scheduling and content
  caching strategy for mobile edge networks using deep reinforcement
  learning,'' in \emph{2018 IEEE International Conference on Communications
  Workshops (ICC Workshops)}.\hskip 1em plus 0.5em minus 0.4em\relax IEEE,
  2018, pp. 1--6.

\bibitem{zhu2018new}
J.~Zhu, Y.~Song, D.~Jiang, and H.~Song, ``A new {deep-Q-learning-based}
  transmission scheduling mechanism for the cognitive internet of things,''
  \emph{IEEE Internet of Things Journal}, vol.~5, no.~4, pp. 2375--2385, 2018.

\bibitem{chu2018reinforcement}
M.~Chu, H.~Li, X.~Liao, and S.~Cui, ``Reinforcement learning based multi-access
  control and battery prediction with energy harvesting in iot systems,''
  \emph{IEEE Internet of Things Journal}, 2018.

\bibitem{kelly1997charging}
F.~Kelly, ``Charging and rate control for elastic traffic,'' \emph{European
  transactions on Telecommunications}, vol.~8, no.~1, pp. 33--37, 1997.

\bibitem{tse2001multiuser}
D.~Tse, ``Multiuser diversity in wireless networks,'' in \emph{Wireless
  Communications Seminar, Standford University}, 2001.

\bibitem{lillicrap2015continuous}
T.~P. Lillicrap, J.~J. Hunt, A.~Pritzel, N.~Heess, T.~Erez, Y.~Tassa,
  D.~Silver, and D.~Wierstra, ``Continuous control with deep reinforcement
  learning,'' \emph{arXiv preprint arXiv:1509.02971}, 2015.

\bibitem{tensorflow2015-whitepaper}
\BIBentryALTinterwordspacing
M.~Abadi \emph{et~al.}, ``{TensorFlow}: Large-scale machine learning on
  heterogeneous systems,'' 2015, software available from tensorflow.org.
  [Online]. Available: \url{https://www.tensorflow.org/}
\BIBentrySTDinterwordspacing

\bibitem{jain1984quantitative}
R.~K. Jain, D.-M.~W. Chiu, and W.~R. Hawe, ``A quantitative measure of fairness
  and discrimination,'' \emph{Eastern Research Laboratory, Digital Equipment
  Corporation, Hudson, MA}, 1984.

\end{thebibliography}
}

\end{document}